\documentstyle[12pt]{article} 
\def\wh{\widehat} 

\def\D{{\cal D}} 

\def\ov{\overline}

\def\NP{{\it Nucl. Phys.}}
\def\CMP{{\it Comm. Math. Phys.}}
\def\PL{{\it Phys. Lett.}} 
\def\PR{{\it Phys.  Rev.}} 
\def\CQG{{\it Class.  Quan.  Grav.}} 
\def\PLMS{{\it Proc.  London Math.  Soc.}}
\def\MRL{{\it Math. Res. Lett.}}
\def\LMP{{\it Lett. Math. Phys.}}
\def\PRp{{\it Phys.  Rep.}}

\begin{document}
\renewcommand{\theequation}{\arabic{section}.\arabic{equation}}

Preprint \hfill {\bf SB/F/98-257} 
\hrule \vskip 1cm
\centerline{\bf ON THE QFT RELATION BETWEEN DONALDSON-WITTEN }
\centerline{\bf INVARIANTS AND FLOER HOMOLOGY THEORY }
\vskip 1cm

\centerline{R.  Gianvittorio and A.Restuccia}
\vskip 4mm 
 
\centerline {\it Universidad Sim\'on Bol\'{\i}var,}
\centerline {\it Departamento de F\'{\i}sica,}
\centerline {\it Apartado Postal 89000, Caracas 1080-A,}
\centerline {\it Venezuela.}
\centerline {\it \ \ \ e-mail: ritagian@usb.ve, arestu@usb.ve}

\vskip 1cm

\centerline {\bf ABSTRACT} 
\vskip .5cm 
A TQFT in terms of general gauge fixing functions is discussed.  
In a covariant gauge it yields the Donaldson-Witten TQFT.  The 
theory is formulated on a generalized phase space where a simplectic structure 
is introduced.  The Hamiltonian is expressed as the anticommutator of 
off-shell nilpotent BRST and anti-BRST charges.  Following original ideas of 
Witten a time reversal operation is introduced and an inner product is 
defined in terms of it.  A non-covariant gauge fixing is presented 
giving rise to a manifestly time reversal invariant Lagrangean and a 
positive definite Hamiltonian, with the inner product previously introduced.  
As a consequence, the indefiniteness problem of some of the kinetic terms 
of the Witten's action is resolved.  The construction allows then 
a consistent interpretation of Floer groups in terms of the 
cohomology of the BRST charge which is explicitly independent of 
the background metric.  The relation between the BRST cohomology and 
the ground states of the Hamiltonian is then completely stablished.
The topological theories arising from the covariant, Donaldson-Witten, and 
non-covariant gauge fixing are shown to be quantum equivalent by 
using the operatorial approach.

\vskip 1cm 
\hrule 
\vfill

\newpage

\section{Introduction}

\setcounter{equation}{0}

Topological Field Theories were introduced by Witten in \cite{W} and have 
had since then a great development from a physical 
and mathematical point of view \cite{W1}-\cite{BG}. Recently an 
interesting relation between the ghost sector of field theories with 
propagating local degrees of freedom and TQFT has been also 
found \cite{BGR}-\cite{S}.

In \cite{W} Witten, inspired on previous work of Atiyah, gave a 
description of the relation of Donaldson and Floer theory.  To define 
Donaldson's invariants of a four manifold with boundary, $X$, one must 
specify a state in the Floer homology of its boundary $\Sigma$ 
\cite{A}.  To do so, on $X$ which Witten assumed to be $\Sigma \times 
\it{R}$, one constructs the topological quantum field theory with effective 
action $S_{eff}$.  One may start with the classical action given in 
\cite{GRS} and construct a canonical Hamiltonian formulation of the 
effective action.  If it is possible to construct a Hilbert space $\cal{H}$ 
of quantum states on $\Sigma$ consistent with a positive Hamiltonian of 
the TQFT, one may then pick a state on $\Sigma$, to construct 
topological invariants of $X$.  These are obtained from path integrals 
with boundary conditions determined by $\psi$:
\begin{equation}
Z= \int \mu \ exp(-S_{eff}) {\cal O} \psi, \label{pi}
\end{equation}
where $\cal O$ is an observable constructed from a product of local 
fields with the property of being BRST invariant:
\begin{equation}
Q {\cal O} =0 ,
\end{equation}
where $Q$ is the off-shell nilpotent BRST operator.
The quantum states $\psi$ must satisfy the physical condition
\begin{equation}
Q \psi =0 .  
\end{equation}
If $\psi$ is Q-exact (\ref{pi}) becomes zero, hence (\ref{pi}) only depends on the 
BRST cohomology class, that is on Q-closed modulo Q-exact fields, which 
are in fact the Floer cohomology classes.  This very relevant relation was 
found in \cite{W}, where the interesting observables $\cal O$ were also 
constructed and related to Donaldson's invariants:
\begin{equation}
{\cal O}=\prod^N_{i=1} \int_{\gamma_i} W_{k_i} ,
\end{equation}
where $W_{k_{i}}$ were constructed from a set of descendent equations 
starting from $ W_0={1\over{2}}Tr{\phi}^2$, $\gamma_{i}$ is a $k_{i}$ 
dimensional homology cycle.  $W_{k}, k=0,\ldots,4$, has ghost number 
$4-k$.  The overall ghost number of $\cal O$ must equal the dimension 
of fermion zero mode space in order to have a non-trivial path 
integral (\ref{pi}).  $\phi$ is the spin zero, BRST invariant field in 
Witten's TQFT.  Following Witten, we are then interested in obtaining a 
canonical Hamiltonian formulation of the TQFT.  In particular of the 
off-shell nilpotent BRST operator $Q$.  The cohomology groups 
associated to $Q$ are precisely the Floer groups.  An explicit 
construction of the metric independent $Q$ should provide also a 
direct proof of the independence on the metric of Floer groups.  
A canonical simplectic formulation of the theory also will help to 
construct a positive definite inner product on the Hilbert space of the 
quantum field theory.  In \cite{W} Witten suggested to use a time reversal 
operation together with a canonical construction in order to consider this 
problem.  The use of a time reversal operation is a natural one in any 
euclidean formulation in quantum field theory \cite{OS}-\cite{M}.  In fact, 
it is essential since it is related to the Wick rotation $t 
\rightarrow it$ because under conjugation $it \rightarrow -it$ while 
$t \rightarrow t$.  In principle, the TQFT does not need any Wick 
rotated Minkowskian formulation so one may think the time reversal is 
not necessarily relevant in this case.  However as suggested by Witten
it turns out to be essential in the definition of the inner product.  

In general it may seem not natural to introduce a hamiltonian 
formulation for a theory over a locally euclidean, orientable manifold 
$X$, since there isn't any natural evolving direction on it.  Moreover 
one usually considers in QFT different boundary conditions on time 
$\pm \infty$ than in any spatial boundary.  In particular a classical 
analysis of a canonical action $\langle p\dot{q}-H(q,p)\rangle$ shows that in 
order to have a gauge invariant canonical action, boundary conditions 
at $t=\pm \infty$ are needed \cite{T}. If we consider an euclidean 
formulation, for example over a 
compact closed  manifold, these boundary conditions have to 
be imposed at some points (corresponding to $t=\pm \infty $).  But 
since there isn't any prefer ``time'' direction on an euclidean manifold 
it seems that the boundary restrictions will appear everywhere.  This 
point was first analysed in \cite{GMR} for the Seiberg-Witten 
topological quantum field theory.  We may argue in the following way.  
We consider the embedding of $X$ into a euclidean manifold $R^n$ of 
high enough dimensions $n$, this is always possible.  We then consider 
a direction in $R^n$ as the euclidean ``time'' $\tau$.  It defines a 
height function over $X$.  One may consider then the hamiltonian 
analysis of the covariant Lagrangean in terms of $\tau$.  Let us call 
A and B the points of lowest and highest height over $X$.  The only 
requirement on the BRST hamiltonian construction with ``time'' $\tau$ 
was found in \cite{BF}-\cite{CR}, it is the quantum analogue of the classical 
conditions on the gauge parameters discussed above:
\begin{equation}
\left.  \left [ Q - \sum _{p} \langle p{\delta Q\over{\delta 
p}}\rangle \right ] \right | ^{\tau_{f}}_{{\tau_{i}}} =0.  \label{pc}
\end{equation}
This condition is satisfied identically by the 
Seiberg-Witten BRST operator \cite{GMR}.  This is a particular 
property of some topological theories.  It will come out that it is 
also satisfied identically by Witten's TQFT.  The canonical 
hamiltonian construction is then completely consistent.  The structure 
of the hamiltonian of Witten's TQFT was obtained in \cite{W}, although 
a canonical simplectic structure was not presented.  It is of the form 
$\{ Q,\bar{Q} \}$, where $Q$ and $\bar{Q}$ were shown to be nilpotent 
on-shell.  This particular form of the hamiltonian was then used, 
under the assumption of the existence of a positive inner product, to 
show that the BRST cohomology consists only of the ground states.  The 
contribution of this paper is to introduce a simplectic structure in 
the formulation of Witten's TQFT, 
to obtain the metric independent off-shell nilpotent BRST charge, 
to construct the time reversal operation generalizing the one introduced 
by Witten and to define an internal product in terms of it.  
We will show then that for the gauge fixing functions we will introduced 
in this work, the Hamiltonian is positive definite resolving the 
problems of indefiniteness of the kinetic terms of Witten's action, 
which are related to the existence of a positive definite inner 
product on the space of quantum states.  The explicit form of the BRST 
charge which is independent of the metric will give a direct proof of 
the topological invariant property of the Floer groups and show that 
consistency condition (\ref{pc}) is identically satisfied.  The BRST 
charge that we will obtain is an extension of the previously found in 
\cite{GRS}.  In that paper the BRST charge was constructed from the 
minimal sector of the extended phase space.  The BRST transformations 
of that sector were there obtained from the Poisson bracket structure.  
The BRST transformations for the non-minimal sector of phase space 
were obtained directly from the general formalism in \cite{BF}-\cite{CR} and 
were non-canonical transformations.  In this paper we obtain a 
simplectic structure which includes the minimal and non-minimal 
sectors.  It allows the canonical construction of $H=\{ Q,\bar{Q} \}$ 
with $Q$ and $\bar{Q}$ related by the time reversal operation and 
off-shell nilpotents.  This structure together with the positive 
definite inner product in the space of states defined in terms of the 
time reversal operation, completes the argument 
in \cite{W} to show that the BRST cohomology consists only on ground 
states.

\section{The gauge invariant action}
\setcounter{equation}{0}

The gauge invariant action introduced in \cite{GRS} is

\begin{equation}
S={1\over{4}}\int Tr(B+F)\wedge (B+F) ={1\over{4}}\int d^4x \;
\varepsilon^{\mu\nu\sigma\rho}(F_{\mu\nu}^a+B_{\mu\nu}^a)
(F_{\sigma\rho}^a+B_{\sigma\rho}^a).  
\label{action}
\end{equation}
 where $F$ is the curvature two form of the gauge connection one 
form $A$, and $B$ is an independent two form.
 
The action (\ref{action}) is invariant under the finite gauge transformation 
on the principal bundle

\begin{eqnarray}
& & A \longrightarrow \Lambda^{-1}A\Lambda+\Lambda^{-1}d\Lambda , \nonumber 
\\
& & B \longrightarrow \Lambda^{-1}B\Lambda , \label{gt}
\end{eqnarray}
and under the infinitesimal gauge transformations with parameter 
$\varepsilon_\mu^a$:

\begin{eqnarray}
&& \delta A_\mu^a =(\D_\mu \omega)^a + 
\varepsilon_\mu^a=\partial_{\mu}\omega^a+(A_\mu \times 
\omega)^a+\varepsilon_\mu^a, \nonumber \\
&&\delta B_{\mu\nu}^a = - (\D_\mu\varepsilon_\nu)^a+ 
(\D_\nu\varepsilon_\mu)^a+(B_{\mu\nu}\times \omega)^a ,  \label{gi}
\end{eqnarray}
where we have also included the infinitesimal transformations 
corresponding to (\ref{gt}). $\varepsilon_\mu^a$ are the gauge parameters 
which eliminate the local degrees of freedom of the theory leaving 
only the topological excitations. We denote $(A_\mu \times
\omega)^a=f^{abc}A^b\omega^c$. The gauge parameter $\varepsilon_\mu^a$ is
restricted by the condition that $A+\delta A$ must also be a connection on the
same principal bundle.
That is, under a finite gauge transformation $\Lambda$ on the principal bundle we have
$$
A+\varepsilon \longrightarrow \Lambda^{-1}(A+\varepsilon)\Lambda
+\Lambda^{-1}d\Lambda 
$$
Consequently
\[
\int Tr F(A+\varepsilon) \wedge F(A+\varepsilon)=\int Tr  F(A) \wedge F(A),
\]
that is the gauge transformations with parameter $\varepsilon$, under the
restriction that $A+\delta A$ is still a connection on the principal bundle, do
not change the Chern Class of F. It can change A within the set of 
connections on the same principal bundle only. 

From now on we supress the group index in order to simplify the expressions. The
gauge invariances (\ref{gi}) allow to make a partial gauge fixing
\begin{equation}
B_{\mu\nu}^{+}=0 
\end{equation}
where

\begin{eqnarray} 
&&B_{\mu\nu}^{\pm} \equiv {1\over{2}} (B_{\mu\nu} \pm 
\ov{B}_{\mu\nu}), \nonumber \\
&&\ov{B}_{\mu\nu} \equiv {\sqrt{g}\over{2}}
\varepsilon_{\mu\nu\sigma\rho}B^{\sigma\rho}.
\end{eqnarray}

In this gauge the field equations reduce to:
\begin{eqnarray}
&&F_{\mu\nu}^{+}=0, \nonumber \\
&&F_{\mu\nu}^{-}+B_{\mu\nu}^{-}=0.  \label{fe}
\end{eqnarray}

This field equations (\ref{fe}) are identical to the ones obtained in 
\cite{W} by Witten.

It is interesting to notice that the action (\ref{action}) give rise by 
dimensional reduction to two dimensions, to a theory decribing as 
field equations the ``self dual" Hitchin equations over Riemann 
surfaces \cite{H}-\cite{MMR}.

We now follow the same steps as in \cite{GRS} but we shall consider a different 
reduction procedure.

The action (\ref{action}) may be reformulated in a canonical form. We 
obtain by direct expansion of the Lagrangean density 
\begin{equation}
S=\int d^4x \ \sqrt{g} \ [ 
\dot{A}_i\varepsilon^{ijk}(F_{jk}+B_{jk})+A_0\D_i[\varepsilon^{ijk}(F_{jk} 
+B_{jk})]+B_{0i}\varepsilon^{ijk}(F_{jk}+B_{jk}) \ ]. 
\end{equation}

Here  we recognize a theory
with vanishing canonical Hamiltonian and with canonical conjugate momenta
to $A_i$ given by 
\begin{equation}
\pi^{i}=\varepsilon^{ijk}(F_{jk}+B_{jk}), 
\end{equation}
where $\varepsilon^{ijk}\equiv \varepsilon^{0ijk}$, and with Lagrange multipliers
$A_0$ and $B_{0i}$ associated respectively to the following constraints

\begin{eqnarray}
&&\phi = \D_i\pi^i=\partial_i\pi^{i}+(A_i\times \pi^i)=0, \nonumber \\
&&\phi^{i}=\pi^{i}=0.  \label{c}
\end{eqnarray}

The algebra of the constraints is given by:
\begin{eqnarray}
&&\{\phi^{ia}(x),\phi^{jb}(x')\} =0, \nonumber \\
&&\{\phi^a(x),\phi^b(x')\} =f^{abc}\phi^c(x)\delta^3(x-x'), \nonumber \\
&&\{\phi^a(x),\phi^{ib}(x')\} =f^{abc}\phi^{ic}(x)\delta^3(x-x'),
\end{eqnarray}
which show that all constraints are first class. Nevertheless they are not
linearly independent since they satisfy the following identity
\begin{equation}
(\D_i\phi^i)=\phi,
\end{equation}
and thus we have to deal with a reducible theory \cite{BF}-\cite{CR} with one level of 
reducibility.  The corresponding matrix of reducibility is given
by: 
\begin{equation} 
a=(\D_i,-1).
\end{equation}

To construct the BRST charge we follow \cite{BF}-\cite{CR} and introduce the minimal
sector of the extended phase space expanded by the conjugate pairs:
\begin{equation} 
(A_i,\pi^{i});(C_1,\mu^{1}),(C_{1i},\mu^{1i});(C_{11},\mu^{11}),
\end{equation} 
where $(A_i,\pi^{i})$ are the original canonical 
coordinates and $(C,\mu)$ are the canonical ghost and antighost 
associated to the constraints (\ref{c}).

The off-shell nilpotent BRST charge associated with (\ref{c}) is then given by:
\begin{eqnarray}
\Omega&=&\langle 
C_1(\D_i\pi^i)+C_{1i}\pi^{i}+C_{11}[(\D_i\mu^{1i})-\mu^{1}] , 
\nonumber\\
& &-{1\over{2}}C_1(C_1\times \mu^{1})-C_1(C_{1i}\times \mu^{1i}) , \nonumber\\ 
& &-C_{11}(C_1\times \mu^{11})\rangle, \label{O}
\end{eqnarray} 
where $\langle \cdots \rangle$ stands for integration on the three 
dimensional continuous index.

We now define the non minimal sector of the
extended phase space \cite{BF}-\cite{CR}. It contains extra ghosts, antighosts and Lagrange
multipliers. First we introduce the C-fields
\begin{equation} 
 C_m,C_{m}^i;\ \ \ C_{mn},C_{mn}^i;\ \ \ \ \ \ m,n=1,2,3
\end{equation} 
where at least one of the indices $m,n$ take the values 
2 or 3.  In addition to these ghost, antighost and Lagrange multiplier 
fields we introduce the $\lambda$ and $\theta$ fields (Lagrange 
multipliers),
also in the non minimal sector, 
\begin{eqnarray}
&&\lambda_1^{0},\ \lambda_{1i}^{0};\ \lambda_{1m}^{0};\ \ m=1,2,3 , 
\nonumber\\
&&\lambda_{11}^{1}; , \nonumber\\
&&\theta_1^{0},\ \theta_{1i}^{0};\ \theta_{1m}^{0};\ \ m=1,2,3 , \nonumber\\ 
&&\theta_{11}^{1}.
\end{eqnarray} 

In this notation the 1 subscripts denote ghost associated to a gauge
symmetry of the action, the 2 subscripts denote antighost associated to a gauge
fixing  condition in the effective action and the 3 subscripts denote
Lagrange  multipliers associated to a gauge fixing condition.
The effective action is then given by:
\begin{eqnarray}
S_{eff}&=&\int d^4x \ \sqrt{g} \ [\pi^{i}\dot{A}_i+\mu^{1}\dot{C}_1+\mu^{1i}\dot{C}_{1i}+
\mu^{11}\dot{C}_{11}  \nonumber \\
& &+\wh{\delta}(\lambda_1^{0}\mu^{1}+\lambda_{1i}^{0}\mu^{1i}+
\lambda_{11}^{1}\mu^{11})  \nonumber\\
& 
&+\wh{\delta}(C_2\chi_2+C_{2}^i\chi_{2i})+\wh{\delta}(C_{12}\chi_{12})] , 
\label{ea}
\end{eqnarray}

In eq.(\ref{ea}) $\chi_2$, $\chi_{2i}$ are the primary gauge fixing 
functions associated to the constraints (\ref{c}) , while $\chi_{12}$, is the 
gauge fixing functions which must fix the longitudinal part of the 
$C_{1i}$ field.  The BRST transformation for the canonical 
variables is given by
\begin{equation}
\wh{\delta}Z=(-1)^{\varepsilon_z}\{ Z,\Omega \},
\end{equation} 
where $\varepsilon_z$ is the grassmanian parity of $Z$. The BRST
transformation of the variables of the non minimal sector are 
determined by imposing the closure of the charge as in \cite{BF}-\cite{CR}.

After the integration on $\mu^{1},\mu^{1i}$ and $\mu^{11}$ we obtain:
\begin{eqnarray}
&&\wh{\delta} \lambda_1^{0}=\D_0C_1+\lambda_{11}^{1}, \nonumber \\
&&\wh{\delta} \lambda_{1i}^{0}=\D_0C_{1i}+\D_i
\lambda_{11}^{1}-\lambda_{1i}^0 \times C_1+{1\over{2}}(C_{2i} \times
C_{11}), \nonumber \\
&&\wh{\delta} \lambda_{11}^{1}=- \D_0C_{11}+(\lambda_{11}^{1} \times C_1),
\end{eqnarray}
where $\D_0 C = \partial_0 C+(A_0 \times C)$ with $A_0=- \lambda_1^0$.
We introduce $C_{1\mu}=(C_{10},C_{1i})$ after we have recognized $C_{10}=-
\lambda_{11}^1=- \lambda_{11}^0$.

After the introduction of the a 
self-dual field $C_{2\mu\nu}$, 
we finally choose gauge fixing functions that may be
written in a covariant form as 
\begin{eqnarray}
&&\chi_2=\partial_{\mu}A^{\mu}-{\alpha\over{2}}C_3,  \nonumber\\
&&\chi_2^{\mu\nu}={a\over{2}}B^{+ \mu\nu}, \ \ \ \ \ a\neq 0,  \nonumber\\
&&\chi_{12}=\D^{\mu}C_{1\mu}+{1\over{2}}
C_{13} \times C_{11}+{1\over{2}}((C_{12}\times C_1)\times C_{11}), \label{gf}
\end{eqnarray}

After elimination of all conjugate momenta the BRST transformation rules
of all the remaining objects take the form 
\begin{eqnarray}
&&\wh{\delta}A_{\mu}=-\D_{\mu}C_1+C_{1\mu},  \nonumber\\
&&\wh{\delta}B^+_{\mu\nu}=-\D_{[\mu}C_{1\nu]}-
{1\over{2}}\varepsilon_{\mu\nu\sigma\rho}
\D^{\sigma}C_1^{\rho}+(C_1\times B^+_{\mu\nu})-
{1\over{2}}(C_{2\mu\nu}\times C_{11}),  \nonumber\\
&&\wh{\delta}C_1=C_{11}+{1\over{2}}(C_1\times C_1),  \nonumber\\
&&\wh{\delta}C_{1\mu}=\D_{\mu}C_{11}+(C_1\times C_{1\mu}),  \nonumber\\
&&\wh{\delta}C_{11}=-(C_{11}\times C_1),  \nonumber\\
&&\wh{\delta}C_2=C_3, \ \ \ \ \ \ \ \ \wh{\delta}C_3=0,  \nonumber\\
&&\wh{\delta}C_{2\mu\nu}=C_{3\mu\nu}=(C_{2\mu\nu}\times
C_1)-2(F^{+}+B^{+})_{\mu\nu}, \nonumber \\
&&\wh{\delta}C_{3\mu\nu}=0, , \nonumber\\ 
&&\wh{\delta}C_{12}=C_{13},  \ \ \ \ \ \ \ \wh{\delta}C_{13}=0. \label{tr}
\end{eqnarray}
 The explicit expression for $C_{3\mu\nu}$ has been obtained
from functional integration. 

It turns out that the algebra  (\ref{tr}) closes off-shell.

The covariant, BRST invariant action, is given by:
 
\begin{eqnarray}
S_{eff}&=&\int d^4x \ \sqrt{g} \left [ {1\over{2}}(F^{+}+B^{+})^2
+\wh{\delta}(C_2 \chi_2) +\wh{\delta}(C_{12} \chi_{12}) \right.\nonumber \\
& & \left. +\wh{\delta}(C_{2\mu\nu} 
\chi_2^{\mu\nu})-{1\over{8}}C_{2\mu\nu}(C_2^{\mu\nu} \times 
C_{11}) \right ],
\label{ea0}
\end{eqnarray}
and it can be rewritten using equations (\ref{gf}) and (\ref{tr}) as follows
\begin{eqnarray}
S_{eff}&=&\int d^4x \ \sqrt{g} \left [ {1\over{2}}F^+_{\mu\nu} 
F^{+\mu\nu}+({1\over{2}}-a)B^+_{\mu\nu} 
B^{+\mu\nu}+(1-a)F^+_{\mu\nu}B^{+\mu\nu} \right.
\nonumber \\
& &+C_2^{\mu\nu}\D_{\mu}C_{1\nu}+{1\over{8}}C_{11}(C_2^{\mu\nu} 
\times C_{2\mu\nu})+\ov{C}_{13}\D_{\mu}C_1^{\mu} \nonumber \\
& &+C_{12}(C_1^\mu \times
C_{1\mu})+C_{12}\D_{\mu}\D^\mu C_{11} +{1\over{2}}C_{11}(\ov{C}_{13}\times 
\ov{C}_{13}) \nonumber \\
& &-{1\over{2}}(C_{12}\times C_{11}) (C_{12}\times 
C_{11})+C_3(\partial_{\mu}A^{\mu}-{\alpha\over{2}}C_3) \nonumber \\ 
& &\left. +C_2\partial_{\mu}\D^\mu C_1-C_2\partial_{\mu}C_1^{\mu} \right ], 
\label{effa}
\end{eqnarray}
with $\ov{C}_{13}=C_{13}+(C_{12}\times C_1)$.  
If we relax (\ref{ea0}) from the gauge fixing $\chi_{2}$ and 
eliminate the auxiliary fields $B_{\mu\nu}$ we
exactly obtain the Witten's TQFT with the following identifications

\begin{eqnarray*}
&&C_{1\mu}=i\psi_{\mu}, \ \ \ C_{11}=-i\phi, \ \ \ \ov{C}_{13}=-\eta,  \\
&&C_{12}={1\over{2}}i\lambda, \ \ \ C_{2\mu\nu}=-\chi_{\mu\nu}.
\end{eqnarray*}

We notice that the sign of the term $B^+B^+$, for any value of $a$
 does not correspond to the one in a Gaussian functional. However 
the $B^+B^+$ term decouples from the action, hence the problem is 
harmless. In fact we may perform a change of 
variable $B^+ \longrightarrow i B^+$ in the functional integral and 
integrate it.  There is also a problem with the sign of the quartic 
term $(C_{12}\times C_{11}) (C_{12}\times C_{11})$, 
since its corresponding kinetic term is indefinite.  
This point was first noticed by Witten in \cite{W}.  The 
functional integral may be correctly defined by eliminating the last 
two terms of the gauge fixing $\chi_{12}$ in eq.(\ref{gf}). In \cite{W} 
$\phi$ and $\lambda$ are taken to be complex conjugates. However this 
implies from (\ref{tr}) that $\wh{\delta}C_1$ becomes complex valued 
and hence inconsistent with the original assumption that the gauge 
parameters $\Lambda$ are real valued. We will resolve this problem by 
following a different approach.

\section{Off-Shell Supersymmetry}
\setcounter{equation}{0}

In order to obtain a gauge supersymmetric action with off-shell closure of the SUSY
algebra, we consider the action

\begin{eqnarray}
S_{eff}&=&\int d^4x \ \sqrt{g} 
\left [ {1\over{2}}(F^+_{\mu\nu} F^{+\mu\nu} - B^+_{\mu\nu} 
B^{+\mu\nu})+C_2^{\mu\nu}\D_{\mu}C_{1\nu}+{1\over{8}}C_{11}(C_2^{\mu\nu}\times 
C_{2\mu\nu}) \right. \nonumber \\
& &+\ov{C}_{13}\D_{\mu}C_1^{\mu}+C_{12}(C_1^\mu \times
C_{1\mu})+C_{12}\D_{\mu}\D^\mu C_{11} \nonumber \\
& &\left. +{1\over{2}}C_{11}(\ov{C}_{13}\times \ov{C}_{13})
-{1\over{2}}(C_{12}\times C_{11}) (C_{12}\times C_{11}) \right ], \label{ea1}
\end{eqnarray}
it arises from (\ref{ea0}) by supressing the gauge fixing $\chi_2$ term. 
The action (\ref{ea1}) is
invariant under the BRST algebra (\ref{tr}). 
It is also invariant under the gauge 
transformations of the $SU(2)$ principal bundle, which are generated 
by the first class constraint $\D_i\pi^i$, since the corresponding 
gauge fixing condition has not been imposed.  Finally (\ref{ea1}) is invariant 
under general coordinate transformations.

If the auxiliarly field $B^+$ is eliminated from (\ref{ea1}) by 
Gaussian integration, we obtain the action proposed by Witten 
\cite{W}
for the TQFT describing Donaldson's invariants as ``topological" observables.

We define the following SUSY transformation $\delta$
$$ 
\delta  \equiv \wh{\delta} \mid_{C_{1}=0}, 
$$
when acting on any of the fields describing the topological theory.

We then have
\begin{eqnarray}
&&\delta A_{\mu}=C_{1\mu}, \nonumber \\
&&\delta \delta A_{\mu}=\delta (\wh{\delta}A\mid_{C_{1}=0}) 
=\wh{\delta} (\wh{\delta}A\mid_{C_{1}=0})\mid_{C_{1}=0} 
=\D_{\mu}C_{11},\nonumber \\
&&\delta C_{1\mu}=\D_{\mu}C_{11}, \nonumber \\
&&\delta \delta C_{1\mu}=C_{1\mu} \times C_{11}, \nonumber \\
&&\delta C_{2\mu\nu}=-2(F^{+}+B^{+})_{\mu\nu}, \nonumber \\
&&\delta \delta C_{2\mu\nu}=C_{2\mu\nu}\times C_{11},\nonumber \\
&&\delta B^+_{\mu\nu}=-\D_{[\mu}C_{1\nu]}-{1\over{2}}\varepsilon_{\mu\nu\sigma\rho}
\D^{\sigma}C_1^\rho -{1\over{2}}(C_{2\mu\nu}\times C_{11}),\nonumber \\
&&\delta \delta B^+_{\mu\nu}=B^+_{\mu\nu}\times C_{11},\nonumber \\
&&\delta C_{11}=0,\nonumber \\
&&\delta C_{12}=C_{13}, \nonumber \\
&&\delta C_{13}=0.
\end{eqnarray}

This algebra closes modulo field dependent gauge transformations (with parameter
$C_{11}$). 

The SUSY transformations defined by Witten are obtained from this 
algebra by eliminating the auxiliarly field $B$. After
this reduction the SUSY algebra closes on-shell and modulo field dependent gauge
transformations. On-shell because the auxiliarly field B has been eliminated, and
modulo field dependent gauge transformations because the Wess-Zumino field, $C_1$ in
this case, has been eliminated.

 In our action (\ref{effa}) we also include 
the Wess-Zumino field $C_1$ allowing the complete off-shell closure of 
the SUSY algebra.  It is interesting to notice that $C_1$ is not the 
usual Wess-Zumino field that is introduced in the superfield 
formulation of the SUSY gauge multiplet.  This is allowed in the 
topological theory because we are only considering the nilpotent 
subalgebra of the full twisted SUSY algebra.

\section{BRST Charge and Hamiltonian of the TQFT on the extended phase space}

\setcounter{equation}{0}

We now introduce an extended phase space in order to define a 
simplectic structure over it.
The canonical conjugate pairs we define are:
\begin{eqnarray}
&&(A_{i}, \pi^i)_{+},\nonumber \\
&&(C_{1}, \mu^1)_{-}, \ (C_{2}, P_{2})_{-}, \ (C_{3}, \lambda^0_{1})_{+}, 
\nonumber \\
&&(C_{1i}, \mu^{1i})_{-}, \ (C_{2}^i, P_{2i})_{-}, \ (C_{3}^i, 
\lambda^0_{1i})_{+}, \nonumber \\
&&(C_{11}, \mu^{11})_{+}, \ (C_{12}, P_{12})_{+}, \ (C_{13}, 
\lambda^1_{11})_{-},\label{cp} 
\end{eqnarray}
where $C_{1}$ and $C_{1i}$ are the ghost fields associated to the 
first class constraints of the theory.  $C_{11}$ arises from the 
reducibility of the first class constraints.  Their ghost numbers are:
\begin{eqnarray*}
&C_{1}&+1\\
&C_{1i}&+1\\ 
&C_{11}&+2\\ 
\end{eqnarray*}
The $C_{2}, C_{2}^i$ and $C_{12}$ fields are associated to the gauge fixing 
conditions and appear in the action as
$$
\wh{\delta}(C_2\chi_2+C_{2}^i\chi_{2i}+C_{12}\chi_{12})  
$$
where $\wh{\delta}$ is the BRST transformation.  $\chi_2$, $\chi_{2i}$ are 
the gauge fixing conditions associated to the first class constraints 
$\phi$ and $\phi^i$ respectively, and have ghost number 0, while 
$\chi_{12}$, is related to the reducibility of $C_{1i}$ and $C_{1}$.  
It has ghost number $+1$.  Consequently the ghost numbers of $C_{2}, 
C_{2}^i$ and $C_{12}$ are:
\begin{eqnarray*}
&C_{2}&-1\\ 
&C_{2}^i&-1\\ 
&C_{12}&-2\\ 
\end{eqnarray*}
The $C_{3}, C_{3}^i$ and $C_{13}$ fields are the Lagrange multipliers 
associated to the gauge fixing conditions.  Their ghost numbers are then:
\begin{eqnarray*}
&C_{3}&0\\ 
&C_{3}^i&0\\ 
&C_{13}&-1\\ 
\end{eqnarray*}
The sum of the ghost numbers of the canonical conjugate pairs must be 
zero, hence the ghost numbers of the other variables in \ref{cp} are 
determined.
The BRST charge in the extended phase space is
\begin{eqnarray}
Q&=&\langle C_1(\D_i\pi^i)+C_{1i}\pi^{i}+C_{11}[(\D_i\mu^{1i})-\mu^{1}] 
\nonumber\\
& &-{1\over{2}}C_1(C_1\times \mu^{1})-C_1(C_{1i}\times \mu^{1i})  \nonumber\\ 
& &-C_{11}(C_1\times \mu^{11})-C_{3}P_{2}-C_{3}^iP_{2i}+C_{13}P_{12}\rangle 
\nonumber \\
&=&\Omega+\langle-C_{3}P_{2}-C_{3}^iP_{2i}+C_{13}P_{12}\rangle.\label{Q}
\end{eqnarray}
The charge Q is off-shell nilpotent
\begin{equation}
\{Q,Q\}=0.
\end{equation}
 The BRST transformation for the variables is given by
\begin{equation}
\wh{\delta}Z=(-1)^{\varepsilon_z}\{ Z,Q\},
\end{equation} 
where $\varepsilon_z$ is the grassmanian parity of $Z$.
 
 In order to express the Hamiltonian of section 3, which reproduces 
 Witten's effective action after functional integration of the 
 auxiliarly fields, as a Q-anticonmutator we introduce the following 
 functional $\tilde{Q}$ which will turn out to be the time reversal of 
 the BRST charge when reduced to the minimal sector of phase space.  
 We define $\tilde{Q}$ in a very general way in terms 
 of general gauge fixing functions, and step by step we impose some 
 consistency conditions over them.  At the end of the process we give 
 the general structure of the gauge fixing functions solving all 
 consistency restrictions.  The $\tilde{Q}$ is given by:
\begin{equation}
\tilde{Q}=\langle C_2\wh\chi_2+C_{2}^i\wh\chi_{2i}+C_{12}\wh\chi_{12}+\lambda_1^{0}\mu^{1} 
+\lambda_{1i}^{0}\mu^{1i}+\lambda_{11}^{1}\mu^{11}\rangle.\label{Qd}
\end{equation}

In order to have off-shell nilpotency of $Q^\dagger$
\begin{equation}
\{\tilde{Q},\tilde{Q}\}=0,
\end{equation}
we impose the following restrictions on the gauge fixing functions:
\begin{equation}
\{\langle C_2\wh\chi_2+C_{2}^i\wh\chi_{2i}+C_{12}\wh\chi_{12}\rangle, 
\langle\lambda_1^{0}\mu^{1} 
+\lambda_{1i}^{0}\mu^{1i}+\lambda_{11}^{1}\mu^{11}\rangle\}=0,
\end{equation}
and
\begin{equation}
\{\langle C_2\wh\chi_2+C_{2}^i\wh\chi_{2i}+C_{12}\wh\chi_{12}\rangle, 
\langle C_2\wh\chi_2 +C_{2}^i\wh\chi_{2i} +C_{12}\wh\chi_{12}\rangle\}=0.
\end{equation}

Later on we will discuss these restrictions, they turn out to be mild 
conditions on the gauge fixing functions.
We then obtain
\begin{eqnarray}
\{Q,\tilde{Q}\}&=&\langle\lambda_1^{0}(\D_i\pi^i-C_{1}\times\mu^1-C_{1i}\times\mu^{1i} 
+C_{11}\times\mu^{11}) \nonumber \\
& &+\lambda_{1i}^{0}(\pi^i-C_{1}\times\mu^{1i}) 
-\lambda_{11}^{1}(\D_i\mu^{1i}-\mu^1-C_{1}\times\mu^{11}) \nonumber \\
& &-C_3\wh\chi_2-C_{3}^i\wh\chi_{2i}-C_{13}\wh\chi_{12}
-P_2\mu^{1}-P_{2i}\mu^{1i}+P_{12}\mu^{11} \nonumber \\
& 
&+C_{2}\{\langle P_2C_3\rangle,\wh\chi_2\}+C_{2}^j\{\langle P_{2i}C_3^i\rangle,
\wh\chi_{2j}\}  \nonumber \\
& &+C_{12}\{\langle P_{12}C_{13}\rangle,\wh\chi_{12}\}-C_{2}\{\Omega,\wh\chi_2\}
\nonumber \\
& & -C_{2}^i\{\Omega,\wh\chi_{2i}\} 
+C_{12}\{\Omega,\wh\chi_{12}\}\rangle.  \label{H}
\end{eqnarray}
The right hand member of eq.(\ref{H}) is exactly the expansion of the terms
\begin{equation}
\langle -\wh{\delta}(\lambda_1^{0}\mu^1+\lambda_{1i}^{0}\mu^{1i}+\lambda_{11}^{1}\mu^{11})
-\wh{\delta}(C_2\wh\chi_2+C_{2}^i\wh\chi_{2i}+C_{12}\wh\chi_{12}) 
\rangle ,
\end{equation}
in eq.(\ref{ea}).  We distinguish $\chi$ from $\wh\chi$ since it is 
more interesting for our formulation to consider
\begin{eqnarray}
&&\chi_2=\wh\chi_2+\dot\lambda_1^{0}, \nonumber \\ 
&&\chi_{2i}=\wh\chi_{2i}+\dot\lambda_{1i}^{0}, \nonumber \\ 
&&\chi_{12}=\wh\chi_{12}-\dot\lambda_{11}^{1}.  \label{gf1}
\end{eqnarray}

We consider in the action (\ref{ea}) the gauge fixing conditions 
(\ref{gf1}).  The time derivative terms contribute to the kinetic part 
of the action which now becomes
\begin{eqnarray}
&&\int d^4x \ \sqrt{g} \ [\pi^{i}\dot{A}_i+\mu^{1}\dot{C}_1+\mu^{1i}\dot{C}_{1i}+ 
\mu^{11}\dot{C}_{11} + 
P_2\dot{C}_2+P_{2i}\dot{C}_2^i+P_{12}\dot{C}_{12} \nonumber\\
&&+\lambda_1^{0}\dot C_{3}+\lambda_{1i}^{0}\dot 
C_{3}^i+\lambda_{11}^{1}\dot C_{13}],
\end{eqnarray}
while the $\wh\chi$ terms contribute to the Hamiltonian of the system 
which has then the expression (\ref{H}).  We have thus expressed the action 
(\ref{ea}) completely in terms of canonical conjugate pairs with the 
Hamiltonian having the form
\begin{equation}
H=\{Q,\tilde{Q}\}, \label{H1}
\end{equation}
with $Q$ and $\tilde{Q}$ off-shell nilpotent charges.
From (\ref{H1}) one obtains
\begin{eqnarray}
&&\{Q, H\}=0, \nonumber \\
&&\{\tilde{Q}, H\}=0.  
\end{eqnarray}

The important remark with respect to the BRST charge (\ref{Q}) is that 
it is independent of the metric over the base manifold $X$.  This is a 
great advantage with respect to the approach in \cite{W}, where the 
BRST charge is obtained from the covariant effective action, already 
metric dependent, and hence the charge is metric dependent.  Our 
canonical approach instead ensures that the BRST charge (\ref{Q}) is 
metric independent.  $\tilde{Q}$  may depend on the metric 
through the gauge fixing functions.  Usually in any field theory the 
gauge fixing conditions leading to a covariant formulation of the 
effective action are metric dependent.  The BRST cohomology, obtained 
from (\ref{Q}), which may be identified to the Floer cohomology is then 
a topological invariant of 
the boundary of $X$.  In \cite{W} it is argued, using (\ref{H1}), how 
to relate the BRST cohomology with the ground states of the 
Hamiltonian.  Let $\psi$ be a state of the quantum theory.  It must 
satisfy
\begin{equation}
Q \psi = 0.
\end{equation}
If $\psi$ is an eigenstate of the Hamiltonian
\begin{equation}
H \psi = \lambda \psi ,
\end{equation}
with $\lambda \ne 0$.  Then
\begin{equation}
\psi = Q \left ( \lambda^{-1} \tilde{Q} \psi \right ) ,
\end{equation}
and this means that $\psi$ is in the trivial cohomology, it is 
$Q-$exact.  Then the cohomology classes of $Q$ correspond to zero 
eigenstate of $H$.  Conversely, if there exists a positive inner product and
\begin{equation}
H \psi = 0 ,
\end{equation}
then
\begin{equation}
 \langle \psi | H | \psi \rangle = {\left |Q |\psi \rangle \right 
 |}^2 + {\left |\tilde{Q} |\psi \rangle \right |}^2 ,
 \label{s}
\end{equation}
and consequently
\begin{eqnarray}
Q |\psi \rangle =0 ,\\
\tilde{Q} |\psi \rangle =0 .
\end{eqnarray}
In (\ref{s}) it is explicitly used the positivity of the inner product.  
However the Hilbert space in \cite{W} as stated by Witten, is indefinite 
because of the terms $\eta \D_{0}\psi_{0}$ and $(\D_{0}\psi_{i}) 
\chi_{i}$ in the Hamiltonian.  Moreover the Hamiltonian in \cite{W} 
is not bounded from below because of the term $\phi \Delta \lambda$ 
when $\phi$ and $\lambda$ are real.  We will discuss these problems in the 
next section.

\section{Time reversal operation and positivity of the Hamiltonian}
\setcounter{equation}{0}

In order to discuss the problems related to the positivity of the 
inner product in the space of quantum states, we will be  
interested in analysing the quadratic part of the 
Hamiltonian, considered as a polinomy on the fields. We will denote 
it $H_{2}$.  It may be expressed as
\begin{equation}
H_{2}=\{Q_{2}, \tilde{Q}_{2} \}, \label{H2}
\end{equation}
where $Q_{2}$ and $\tilde{Q}_{2}$ are the quadratic parts, when 
considered as a polinomy on the fields, of the 
BRST and anti-BRST charges:
\begin{eqnarray}
&&\{Q_{2},Q_{2}\}=0, \\
&&\{\tilde{Q}_{2},\tilde{Q}_{2} \}=0.
\end{eqnarray}
We may now introduce the time reversal operation $\it T$:  
\renewcommand{\arraystretch}{1.5}
\begin{equation}
\begin{array}{rcl}
P_{2}&\stackrel{\it T}{\longrightarrow} &C_{2} \\
P_{2i}&\stackrel{\it T}{\longrightarrow} &C_{2}^i\\
P_{12}&\stackrel{\it T}{\longrightarrow} &C_{12} \\
C_{3}&\stackrel{\it T}{\longrightarrow} &-\wh\chi_2\\
C_{3}^i&\stackrel{\it T}{\longrightarrow} &-\wh\chi_{2i}\\
C_{13}&\stackrel{\it T}{\longrightarrow} &\wh\chi_{12} \\
C_{11}&\stackrel{\it T}{\longrightarrow} &\mu^{11}\\
(\partial_i\mu^{1i}-\mu^{1})&\stackrel{\it T}{\longrightarrow} 
&\lambda_{11}^{1} \\
\pi^i&\stackrel{\it T}{\longrightarrow} &\lambda_{1i}^{0} \\
C_{1}&\stackrel{\it T}{\longrightarrow} &\mu^{1}\\
C_{1i} &\stackrel{\it T}{\longrightarrow}&\mu^{1i} \\
\end{array}
\label{dt}
\end{equation}
where $\partial_{i}$ denotes the 
covariant derivative with respect to the background metric.  
The time reversal operation T by definition satisfies ${\it T}^2=1$. 
This implies that 
\begin{equation}
\lambda_{11}^{1}=\partial^{i}C_{1i}-C_{1} ,
\end{equation}
which will arise from the gauge fixing procedure.  
For the BRST charge we have
\begin{equation}
Q_{2}\stackrel{\it 
T}{\longrightarrow}\tilde{Q}_{2}\stackrel{\it 
T}{\longrightarrow}Q_{2} \ ,
\end{equation}

We will denote $\phi^\dagger \equiv T \phi$. Later on we will 
interpret $\dagger$ as the adjoint under the internal product we will 
introduce.

We will consider now  several canonical reductions of the topological 
action.  All of them will be performed on the full action and then 
discuss properties of the quadratic part of the Hamiltonian. We first assume
\begin{eqnarray}
&&\{\langle P_{2}C_3\rangle,\chi_{2}\}=a P_{2},\nonumber \\
&&\{\langle P_{2i}C_3^i\rangle,\chi_{2j}\}=b P_{2j} ,\nonumber \\
&&\{\langle P_{12}C_{13}\rangle,\chi_{12}\}=d P_{12},
\end{eqnarray} 
where $a, b, d$ are real numbers to be determined.
We then functionally integrate on $P_{2}, P_{2i}, P_{12}$.  We obtain
\begin{equation}
\delta(\mu^{1}+aC_{2}) \ \delta(\mu^{1i}+bC_{2}^i) \ 
\delta(\mu^{11}+dC_{12}) , \label{delta}
\end{equation}
in the functional measure.  We finally integrate on $C_{2}, C_{2}^i, 
C_{12}$ and obtain the Hamiltonian in the reduced phase space:
\begin{eqnarray}
H&=&\langle\lambda_1^{0}(\D_i\pi^i-C_{1}\times\mu^1-C_{1i}\times\mu^{1i} 
+C_{11}\times\mu^{11}) \nonumber \\
& &+\lambda_{1i}^{0}(\pi^i-C_{1}\times\mu^{1i}) 
-\lambda_{11}^{1}(\D_i\mu^{1i}-\mu^1-C_{1}\times\mu^{11}) \nonumber \\
& &-C_3\wh\chi_2-C_{3}^i\wh\chi_{2i}-C_{13}\wh\chi_{12}
+ {1\over{a}}\mu^1\{\Omega,\wh\chi_2\}\nonumber \\
 & & +{1\over{b}}\mu^{1i}\{\Omega,\wh\chi_{2i}\} 
 -{1\over{d}}\mu^{11}\{\Omega,\wh\chi_{12}\}\rangle.  \label{HH}
\end{eqnarray}
 
We will impose further conditions on the gauge fixing functions.  They 
are 
\begin{eqnarray}
&&\{\Omega ,\chi_{2}\}=-\alpha (\Delta C_{1} - \partial^{i}C_{1i}),\nonumber \\
&&\{\Omega ,\chi_{2i}\}=-\eta ( C_{1i} - \partial_{i}C_{1}),\nonumber \\
&&\{\Omega ,\chi_{12}\}=\gamma (\Delta C_{11}-C_{11}),
\end{eqnarray} 
where 
\begin{eqnarray}
&&\frac{\gamma} {d} > 0 , \nonumber \\
&&{\alpha \over{a}}=-{\eta \over{b}} > 0 . \label{restr}
\end{eqnarray}
We have then imposed several restrictions to the gauge fixing 
functions.  It turns out that a solution to all of them is given by
\begin{eqnarray}
&&\chi_2=a\lambda_1^{0}-\alpha\partial^{i}A_{i}, \nonumber \\
&&\chi_{2i}=b\lambda_{1i}^{0}+ \eta A_{i}, \nonumber \\
&&\chi_{12}=d\lambda_{11}^{1}-\gamma(\partial^{i}C_{1i}-C_{1}).  \label{gf2}
\end{eqnarray}
These gauge fixing functions may be continuously deformed with linear 
and nonlinear terms.  They are all admisible gauge fixing functions.  
The QFT should be independent of the element in the admissible set.  
This is though if there isn't a gauge anomaly in the theory. We will 
consider this point in the next section.

We may now perform a further canonical reduction.  We will eliminate the 
pair of conjugate variables $(C_{3}, \lambda^0_{1}), (C_{3}^i, 
\lambda^0_{1i})$ and $(C_{13}, \lambda^1_{11})$ which appear linearly 
in the full action. In order to perform the reduction we integrate 
on $C_{3}, C_{3}^i, C_{13}$.  We obtain in the functional measure
\begin{equation}
\delta(\chi_{2}) \ \delta(\chi_{2i}) \ \delta(\chi_{12}) .
\end{equation}
We may now integrate on $ \lambda^0_{1}, \lambda^0_{1i}$ and 
$\lambda^1_{11}$. 
After this final canonical reduction we have returned to the minimal 
sector of phase space.  In the process, however, we have been able to 
find gauge fixing functions which yield a canonical action whose 
quadratic part is invariant under 
$\it T$ and with a positive definite quadratic part of the Hamiltonian.  
The explicit expressions for the quadratic parts are
\begin{eqnarray}
S_{2}&=&\int d\tau \ [\langle \pi^{i}\dot{A}_i+\mu^{1}\dot{C}_1+\mu^{1i}\dot{C}_{1i}+ 
\mu^{11}\dot{C}_{11} \rangle - H_{2}]  , \\
H_{2}&=& \langle \pi^{i\dagger} \pi^i + \frac{\alpha}{a} 
(\partial_i\pi^{i})^{\dagger} \partial_i\pi^{i}
+\frac{\gamma}{d}(\partial^{i}C_{1i}-C_{1})^\dagger
(\partial^{i}C_{1i}-C_{1}) \nonumber \\
 & &+\frac{\alpha}{a} (C_{1i}-\partial_{i}C_{1})^\dagger 
 (C_{1i}-\partial_{i}C_{1}) 
 +\frac{\gamma}{d}(\partial_{i}C_{11})^\dagger (\partial_{i}C_{11}) 
 \nonumber \\
& &+\frac{\gamma}{d} C_{11}^\dagger C_{11} \rangle.
 \label{H222}
\end{eqnarray}
After the elimination of $\lambda^0_{1i}$ we obtain from (\ref{dt})
\begin{equation}
T \pi^{i}=\lambda^0_{1i}=A_{i} , \label{dt2}
\end{equation}
and since $T^{2}=1$
\begin{equation}
T A_{i}=\pi^{i} , \label{dt3}
\end{equation}
we have completed the table (\ref{dt}) for all canonical conjugate 
pairs.
It is important to remark several points with respect to the structure 
of $H_{2}$ in (\ref{H222}).  In \cite{W} the fields $\phi$ and $\lambda$ 
were taken to be complex and satisfying
\begin{equation}
\phi= - \lambda^{\star} ,
\end{equation}
in order to ensure the positivity of the $(\phi, \lambda)$ kinetic terms in 
the action of the TQFT.  The drawback of this approach, as explained 
by Witten, is that the Lagrangian is then not real.  Consequently the 
quantum field correlation function giving rise to the Donaldson 
invariants are not manifestly real.  In our approach, $C_{11}$ and 
$\mu^{11}$ are real fields related by the time reversal operation. In 
the reduction procedure we obtain from (\ref{delta})  $C_{12}$ in 
terms of $\mu^{11}$.
The corresponding quadratic 
term is then manifestly positive definite when one defines the inner 
product on the Hilbert space of physical states $\cal{H}$ as
\begin{equation}
(\psi,\varphi)_{+}\equiv (\it T \psi,\varphi)  \ , \label{ip}
\end{equation}
where $(\ ,\ )$ is the $L^2$ inner product.  The advantage of the 
approach we have followed as suggested in \cite{W} is that the adjoint in the sense of 
$(\ ,\ )_{+}$ of $\phi$ is $\lambda$ and viceversa, while in terms of the 
$(\ ,\ )$ inner product they are self-adjoints.  The same happens to 
$Q$ and $Q^{\dagger}$, which are adjoint under the $(\ ,\ )_{+}$ 
inner product and this is precisely the property needed in the 
argument used to show that all ground states of the Hamiltonian 
are physical states.
The next point to emphasize is with respect to the indefinite terms 
$\eta \D_{0} \psi$ and $(\D_{0} \psi_{i}) \chi_{i}$ appearing in 
\cite{W}.  They authomatically yield, for self-adjoint fields, an 
indefinite Hilbert space inner product.  In fact,
\begin{eqnarray}
&&\{ \psi (\sigma), \eta (\sigma^{'}) \} = \delta_{\sigma \sigma^{'}} , \nonumber \\
&&\{ \psi^i (\sigma), \chi_{j} (\sigma^{'}) \} =\delta^i_{j} \delta_{\sigma 
\sigma^{'}} \ ,  \label{ip2}
\end{eqnarray}
and all the others anticommutators are zero.

We then applied a bra and a ket from the right and left respectively 
to all anticommutators and use the 
self-adjoint property of the fields to obtain from the left member a 
positive definite matrix while from the right an indefinite matrix, 
that is an inconsistency unless 
the inner product is not positive definite.  In our approach, the 
definition (\ref{ip}) , allows to obtain positive definite results 
from the right and left members of (\ref{ip2}).  This is though 
because under the time reversal operation:
\begin{equation}
\begin{array}{rcccl}
\psi&\stackrel{\it T}{\longrightarrow} &\eta&\stackrel{\it 
T}{\longrightarrow}&\psi \\
\chi_{i}&\stackrel{\it T}{\longrightarrow} &\psi_{i}&\stackrel{\it 
T}{\longrightarrow}&\chi_{i} \ .
\end{array}
\end{equation}
We then have
\begin{eqnarray}
&&(\varphi, \psi\eta\varphi)_{+} = (\it T \varphi,\psi\eta\varphi) =
 (\psi \it T \varphi,\eta\varphi) = \nonumber \\
&&(\it T \it T \psi \it T \varphi,\eta\varphi) = 
(\it T(\eta \varphi),\eta\varphi) =
(\eta \varphi,\eta\varphi)_{+} \ ,  \nonumber 
\end{eqnarray}
that is, the matrix from the right hand side in the previous argument 
is now also positive definite.

We have thus obtained a canonical version of Witten's TQFT with a 
different set of gauge fixing functions.  Starting from the classical action 
(\ref{action}) we have performed a canonical construction of the 
effective action on general gauge fixing functions.  In the covariant 
gauge (section 2) we recover Witten's TQFT effective action with 
manifest covariance.  In the new set of gauge fixing functions 
defined in this paper (\ref{gf2}) we obtain an effective action whose 
quadratic part is invariant under time reversal, 
with a positive definite Hamiltonian 
$H_{2}$ in terms of the inner product introduced in (\ref{ip}). The 
necessary conditions to have a Hamiltonian consistent with a positive 
inner product, raised in \cite{W}, are then satisfied.
The interpretation of the Floer's theory in terms of a quantum field 
theory may then be performed with the gauge fixing functions 
introduced in this section. 

\section{Effective Action of the TQFT}
\setcounter{equation}{0}
We will present in this section an admissible deformation of 
(\ref{gf2}) with non-linear terms which allows to rewrite the 
Hamiltonian (\ref{HH}) in a manifestly positive form. It can be 
deduced from (\ref{HH}) that a convenient non-linear admissible 
deformation of (\ref{gf2}) is given by
\begin{eqnarray}
&&\chi_2=a\lambda_1^{0}-\alpha(\D_i\pi^{i}-\frac{1}{2} C_{1}\times \mu^{1})^{\dagger}, \nonumber \\
&&\chi_{2i}=b\lambda_{1i}^{0}+\eta (\pi^{i}-C_{1}\times 
\mu^{1i})^{\dagger}, \nonumber \\
&&\chi_{12}=d\lambda_{11}^{1}-\gamma(\D_i\mu^{1i}-\mu^{1}
-C_{1}\times \mu^{11})^{\dagger}.  \label{cgf2}
\end{eqnarray}
We consider in what follows
\begin{eqnarray*}
&&\alpha=a=1 \\
&&\eta=-b=-1 \\
&&\gamma=d=1
\end{eqnarray*}
which satisfy the restriction (\ref{restr}).
Using (\ref{dt}) and (\ref{dt2}-\ref{dt3}) we obtain
\begin{eqnarray}
&&\chi_2=\lambda_1^{0}-(\partial^{i}A_{i}-A_{i} \times \pi^{i}+
\frac{1}{2} C_{1}\times \mu^{1}), \nonumber \\
&&\chi_{2i}=\lambda_{1i}^{0}-(A_{i}+C_{1i}\times 
\mu^{1}), \nonumber \\
&&\chi_{12}=\lambda_{11}^{1}-(\partial^{i}C_{1i}+\pi^{i} \times C_{1i}
-C_{1}+C_{11}\times \mu^{1}),  \label{cgf3}
\end{eqnarray}
where we have used the convention
\[
(AB)^{\dagger}=B^{\dagger}A^{\dagger}.
\]

We may now insert (\ref{cgf3}) into (\ref{HH}) and perform the same 
canonical reduction as before. We eliminate then the canonical 
conjugate pairs $(C_{3}, \lambda^0_{1}), (C_{3}^i, \lambda^0_{1i})$ 
and $(C_{13}, \lambda^1_{11})$.

After this canonical reduction we end with a description of the theory 
in terms of the minimal sector of phase space as in section 5 but 
now for the complete effective action. After several calculations we 
obtain
\begin{eqnarray}
S&=& \int d\tau \ [\langle \pi^{i}\dot{A}_i+\mu^{1}\dot{C}_1+\mu^{1i}\dot{C}_{1i}+
\mu^{11}\dot{C}_{11}\rangle-H] \nonumber \\
H&=& \langle(\pi^{i}-C_{1}\times \mu^{1i})^{\dagger} 
(\pi^{i}-C_{1}\times \mu^{1i})  \nonumber \\
& &+  (\D_i\pi^i-C_{1}\times \mu^{1}-C_{1i}\times \mu^{1i}+C_{11}\times 
\mu^{11})^{\dagger} 
(\D_i\pi^i-C_{1}\times \mu^{1} \nonumber \\
& & -C_{1i}\times \mu^{1i}+C_{11}\times \mu^{11}) 
+ (C_{1i}-\D_iC_{1})^\dagger (C_{1i}-\D_i C_{1}) \nonumber \\
& &+ (\partial^{i}C_{1i}-C_{1}+\pi^{i}\times C_{1i}-\mu^{1}\times C_{11})^\dagger
(\partial^{i}C_{1i}-C_{1}+\pi^{i}\times C_{1i}-\mu^{1}\times C_{11}) 
\nonumber \\
& &+(\mu^{11} \times C_{1i}+\mu^{1} \times A_{i})^{\dagger}
(\mu^{11} \times C_{1i}+\mu^{1} \times A_{i})\nonumber \\
& &+(\D_i C_{11}+C_{1}\times C_{1i})^\dagger (\D_i C_{11}+C_{1}\times 
C_{1i}) \nonumber \\
& &+(C_{11}+\frac{1}{2} C_{1}\times C_{1})^\dagger (C_{11}+\frac{1}{2} C_{1}\times C_{1})
 \nonumber \\
& &+ (C_{1}\times C_{1i})^\dagger (C_{1}\times C_{1i})
+(C_{11}\times C_{1})^\dagger (C_{11}\times C_{1}) \rangle.
 \label{HRnl}
\end{eqnarray}

The Hamiltonian (\ref{HRnl}) may be expressed as the Poisson bracket 
of the off-shell nilpotent BRST charge $\Omega$ (\ref{O}), and 
anti-BRST charge $\Omega^{\dagger}$:
\begin{eqnarray}
\Omega ^\dagger = T \Omega \nonumber \\
\{\Omega, \Omega \}=0, \nonumber \\
\{\Omega^{\dagger}, \Omega ^{\dagger} \}=0, \nonumber \\
H=\{\Omega, \Omega ^\dagger \}, \label{HR}
\end{eqnarray}

We notice that the complete effective action (\ref{HRnl}) is 
invariant under the time reversal operation.
We have constructed then gauge fixing functions (\ref{cgf2}), which 
are admissible deformations of (\ref{gf2}), yielding an effective 
BRST invariant action, consistent with a positive inner product in the 
space of Hilbert states.
The quantum equivalence between the effective action (\ref{HRnl}) and Witten's 
effective action arises from the independence of the functional 
integral on the gauge fixing functions.  This latest point is based on 
the BRST invariance of the canonical action \cite{BF}-\cite{CR}.  In order to 
ensure this point, the requirement of nilpotency of $Q$ has to be 
raised, to the quantum level.  That is
\begin{equation}
\{Q,Q\}=0, \label{np1}
\end{equation}
as an operatorial condition.
 
Given an ordering for the BRST charge $Q$ then the ordering for $H$ is 
authomatically determined from
\begin{equation}
H=\{Q,Q^\dagger\}.
\end{equation}
The expression (\ref{Q}) of the BRST charge is given in the $qp$ order 
and as already said is classically nilpotent off-shell.  We consider 
now the situation when $Q$ is an operator constructed from the 
canonical quantization approach.  It has the property that is linear 
in the conjugate momenta variable.  Consequently for any conmutator of 
the form
\begin{equation}
\left [ \prod _{L} q \, p_{j} \, , \prod_{K} \hat{q} \, \hat{p}_{i} \right ]
\label{C}
\end{equation}
where $ \prod q$ denotes a product of $q_{l}$ operators for $l \in L$ and 
$ \prod \hat {q}$ another product of $\hat{q}$ operators at a different 
point, and the conjugate pairs being $(q_{i} \, , p_{i}) \; i=1, 
\ldots ,n$, we expand (\ref{C}) and obtain
\begin{equation}
\prod_{K} \hat{q} \left [ \prod_{L} q \, , \hat{p}_{i} \right ]p_{j}+ 
\prod _{L} q \left [p_{j} \, , \prod _{K} \hat{q} \right ]\hat{p}_{i}
\label{CC}
\end{equation}
which is again linear in the momenta and also $qp$ ordered.
Any cancellation that was valid classically using Poisson brackets is 
then valid at the operatorial level.  In fact the only differene between 
the two evaluations is in the ordering of the resulting terms in 
(\ref{CC}).  Two polinomic terms which cancel classically may not do it as 
operators because of possible different orderings of non conmuting 
operators in the polinomy.  However in the TQFT under consideration 
the resulting expression (\ref{CC}) has a determined $qp$ ordering and 
consequently any classical cancellation ensures an operatorial one.  It 
can then be shown that the expression for the BRST Charge $Q$ in 
(\ref{Q}) satisfies (\ref{np1}) as an operator.  This important 
property ensures that there are no gauge anomalies in this QFT and 
hence the quantum equivalence of Witten's TQFT effective action and 
the one presented in this paper is assured.  Having defined $H=\{ Q, 
Q^{\dagger} \}$, the nilpotency of $Q$ ensures the operatorial relation
 \begin{equation}
 \left [ H,Q \right ]=0
 \end{equation}
 The property of linearity of the BRST charge $Q$ in the 
 momenta may also be used as in \cite{CGRS} to show that the TQFT is 
 independent of the coupling constant.  This property is directly 
 related to the independence of the partition function on the 
 background metric as explained in \cite{W}.  Finally the linearitity 
 of $Q$ also ensures the consistency of the Hamiltonian approach over a 
 compact euclidean orientable manifold $M$ \cite{GMR}.  The point is 
 that one may always embed $M$ over $R^N$ for $N$ large enough.  One 
 can then consider a height function over $M$ by taking a direction on 
 $R^N$.  This defines an euclidean time over $M$, and allows to follow 
 a canonical approach as we have done.  The consistency requirement 
 arises from the BRST invariance of the effective action.  It requires 
 boundary conditions at the points of heighest and lowest $\tau$.  
 Since we may choose almost any direction on $R^N$ to define $\tau$, we 
 end up with consistency conditions almost every where over $M$.  
 Fortunately when $Q$ is linear in the momenta, the boundary condition 
 is satisfied identically giving rise to a consistent Hamiltonian 
 approach.  The boundary conditions on other fields theories with 
 locally propagating degrees of freedom are in general non trivial and 
 with a relevant physical intpretation \cite{BF}-\cite{CR}.

\section{Conclusions}
\setcounter{equation}{0}

We obtained the canonical structure for Witten's TQFT allowing the 
description of the Floer theory in terms of a Hamiltonian consistent 
with a positive inner product.  
We started from a gauge action introduced in \cite{GRS} 
and by considering a covariant gauge fixing BRST procedure we obtained 
Witten's effective action including auxiliarly fields.  We then 
found an extension of the phase space, where the explicit 
expression of the off-shell nilpotent charge is obtained.  The 
Hamiltonian is expressed in the form $\{Q,Q^\dagger \}$ where $Q$ and
$Q^\dagger$ are the BRST and anti-BRST nilpotent charge.  The $Q^\dagger$ is 
expressed in a general form in terms of gauge fixing functions 
satisfying necessary requirements to obtain nilpotency of the 
anti-BRST charge.  The explicit expression of $Q$ is metric 
independent, the resulting BRST cohomology is then identified, 
following Witten, to the Floer groups.  This gives a direct proof that 
Floer's groups are topological invariants depending on the boundary of 
the base manifold $X$.  The Hamiltonian approach is shown to be 
consistently defined by checking that the boundary conditions arising 
from the BRST construction are identically satisfied.  The time 
reversal operation introduced by Witten in \cite{W} is generalized to 
the extended phase space.  It is shown that the necessary conditions 
raised in \cite{W}  to 
have a positive definite inner product in the Hilbert space of states 
are satisfied provided that suitable admissible gauge fixing functions are 
choosen.  The resulting TQFT with those gauge fixing functions is not 
manifestly covariant but satisfies the positivity requirement.  
Finally by going to the operatorial formulation it is shown that the 
nilpotency condition on the BRST operator is satisfied as a quantum 
operator.  This property 
ensures that Witten's TQFT, manifestly covariant, and the one obtained 
from the gauge fixing functions introduced in this paper are quantum 
equivalent field theories.  The two properties, covariance and 
positivity are then obtained by considering different gauge fixing 
functions of the same TQFT, arising from the classical gauge action 
introduced in \cite{GRS}.

\section{Acknowledgement} We would like to thank Prof.  M.F.  Atiyah 
for illuminating comments on TQF theories.  R.G.  would like to thank 
the HEP Group  at the University of Pennsylvannia and in particular 
Prof. Burt Ovrut for helpful discussions. We are also grateful to 
Prof. Jorge Stephany for valuable comments.

\end{document}